\documentclass[twocolumn,showpacs,prb]{revtex4}
\usepackage{amsfonts}
\usepackage{amsmath}
\usepackage{amssymb}
\usepackage{graphicx}

\input{tcilatex}
\begin{document}

\preprint{}
\title{Effective tight-binding model for the iron vacancy ordered K$_{y}$Fe$%
_{1.6}$Se$_{2}$ }
\author{Shin-Ming Huang$^{1}$ and Chung-Yu Mou$^{1,2,3}$}
\affiliation{$^1$Department of Physics, National Tsing Hua University, Hsinchu 30043,
Taiwan\\
$^2$Institute of Physics, Academia Sinica, Nankang, Taiwan\\
$^3$Physics Division, National Center for Theoretical Sciences, P.O.Box
2-131, Hsinchu, Taiwan}

\begin{abstract}
We investigate the electronic structure of the ternary iron selenide K$_{y}$%
Fe$_{1.6}$Se$_{2}$ by considering the spatial symmetry of the $\sqrt{5}%
\times \sqrt{5}$ vacancy ordered structure. Based on three orbitals of $%
t_{2g}$, which are believed to play major physics in iron-based
superconductors, an effective two-dimensional tight binding Hamiltonian is
constructed with the vacancy ordered structure being explicitly included. It
is shown that the constructed band model, when combined with generalized
Hubbard interactions, yields a spin susceptibility which exhibits both the
block-checkerboard antiferromagnetism instability and the stripe
antiferromagnetism instability. In particular, for large Hund's rule
couplings, the block-checkerboard antiferromagnetism wins over the stripe
antiferromagnetism, in agreement with the observation in experiments. We
argue that such a model with correct symmetry and Fermi surface structures
should be the starting point to model K$_{y}$Fe$_{1.6}$Se$_{2}$. The spin
fluctuations at $\mathbf{q}$=($\pi ,\pi $) suggest that interblock
fluctuations of spins might play an important role in the mechanism of
superconductivity occurring in this system.
\end{abstract}

\maketitle

\section{Introduction}

After the discovery of superconductivity in LaFeAsO \cite{Kamihara}, the
finding of iron-chalcogenide superconductor $\alpha $-FeSe$_{x}$ \cite{FCHsu}
has stimulated another intensive studies on the iron-based superconductors.
Although the iron-chalcogenides and iron-pnictides have similar crystal and
electronic structures, they show many differences either in
superconductivity or in magnetism. For instance, the former has lower $T_{c}$
($\sim 8K$) \cite{FCHsu}\ and a larger magnetic moment of the Fe ion in FeTe$%
_{1-x}$Se$_{x}$ ($\sim 2\mu _{B}$), while the latter, \textit{e.g. }La-1111,
has much higher $T_{c}=26K$ \cite{Kamihara}\ but has a smaller magnetic
moment ($\sim 0.36\mu _{B}$) for the Fe ion\cite{Cruz}. Even the magnetic
orders are different, one is bi-collinear and the other is collinear.
Recently, K, Cs or Rb intercalated FeSe superconductors A$_{y}$Fe$_{2-x}$Se$%
_{2}$ are found. It is shown that $T_{c}$ of this system can be enhanced
above $30K$ \cite{JGuo,Maziopa,AFWang}. In this system, as the atomic ratio
of Fe:Se is not 1:1, which used to be in iron-chalcogenides, iron deficiency
is produced. As a result, in addition to the enhancement of $T_{c}$, a $%
\sqrt{5}\times \sqrt{5}$\ iron vacancy ordered pattern illustrated in Fig. %
\ref{lattice_245} is formed when the composition is close to K$_{y}$Fe$_{1.6}
$Se$_{2}$ \cite{WBao,Bacsa,ZWang}. The vacancy ordering is followed by the
magnetic transition at lower temperature $T_{N}=560K$ to the block
antiferromagnetic (AFM) phase with a big moment 3.31$\mu _{B}$ \cite{WBao}.
Although the AFM state is observed in superconducting K$_{0.8}$Fe$_{1.6}$Se$%
_{2}$, evidence of nanoscale phase separation from X-ray diffraction is
reported \cite{Ricci}.

Previous theoretical works \cite{TAMaier,FaWang,YZhou,TDas1,Mazin} on
superconductivity in the A$_{y}$Fe$_{1.6}$Se$_{2}$ system were based on the
band structure of KFe$_{2}$Se$_{2}$ in which the hole pocket around $\Gamma $%
\ is absent and only electron pockets are present at M. Such Fermi surface
tomography was supported by ARPES \cite{YZhang,DMou,TQian,LZhao}. However,
since there is no experimental evidence \cite{YZhang,TQian} showing the
existence of the $\sqrt{5}\times \sqrt{5}$ pattern in KFe$_{2}$Se$_{2}$, the
validity of these approaches is questionable. In fact, because of the $\sqrt{%
5}\times \sqrt{5}$ pattern, the symmetry group changes from \textit{I4/mmm}
to \textit{I4/m} and both the unit cell and the Brillouin zone (BZ) change
as well. The underlying band structure should be very different. Indeed, the
first-principles calculations of the $\sqrt{5}\times \sqrt{5}$ vacancy
ordered lattice structure \cite{XWYan,Cao} indicate that a hole pocket at $%
\Gamma $\ appears in the nonmagnetic state and as expected it loses the
reflection symmetry in the \textit{x-y} plane. The presence of a hole pocket
at $\Gamma $ point indicates that the physics that drives superconductivity
could be very different. It thus calls for a close examination based on an
appropriate Hamiltonian to model the ternary iron selenide K$_{y}$Fe$_{1.6}$%
Se$_{2}$.

In this paper, we construct a two-dimensional tight-binging model with three 
$t_{2g}$ orbitals for the K$_{0.8}$Fe$_{1.6}$Se$_{2}$\ system with the $%
\sqrt{5}\times \sqrt{5}$ vacancy ordered pattern being explicitly included.
Based on the general tight-binding model $H_{t}$ with symmetry imposed by
the vacancy order, we fit the dispersion relation of $H_{t}$ to that of the
non-magnetic state from the first-principles calculations of Ref. \cite%
{XWYan}. Two hole pockets and two electron pockets emerges in the fitted
tight-binding model. The constructed band model, when combined with
generalized Hubbard interactions, yields a spin susceptibility that shows a
large peak around the $\mathbf{q}$-vector ($\pi ,\pi $) for the
block-checkerboard AFM state. Furthermore, competition of different magnetic
states is found but the block-checkerboard AFM state gets enhanced with
larger Hund's rule coupling and wins out at the end. The implication of our
results to the mechanism of superconductivity is discussed. In particular,
we argue that the spin fluctuations at $\mathbf{q}$=($\pi ,\pi $) suggest
that in analogy to spin-fluctuations in high $T_{c}$ cuprates, the
interblock fluctuations of spins might play an important role in the
mechanism of superconductivity occurring in this system.

\section{Theoretical Model}

\begin{figure}[tbp]
\begin{center}
\includegraphics[height=1.7988in, width=2.5469in ] {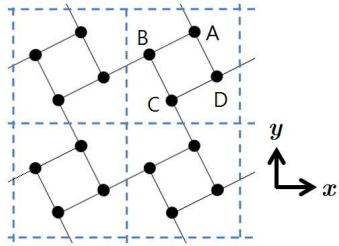}
\end{center}
\caption{(Color online) The schematic representation of the $\protect\sqrt{5}%
\times \protect\sqrt{5}$ vacancy ordered lattice structure of the iron
plane. The black dots denote Fe atoms and the dashed lines enclose the unit
cells. $x$ and $y$ are the primitive vectors.}
\label{lattice_245}
\end{figure}

We start by considering the vacancy-ordered structure with right-handed
chirality \cite{XWYan} as shown in Fig. \ref{lattice_245}. In this ordered
state, the unit cell includes four iron atoms (without considering Se),
which we denote as $I=A$, $B$, $C$, and $D$ and three\textbf{\ }$t_{2g}$%
\textbf{\ }orbitals ($d_{\overline{x}\overline{z}}$, $d_{\overline{y}%
\overline{z}}$, and $d_{\overline{x}\overline{y}}$) are considered in each
iron. Therefore we have 12 species of electrons in a unit cell. We will
suppress spin indices and denote the electron operators collectively as a
vector by $\psi =\left( \psi _{1},\psi _{2},\psi _{3}\right) $ with $\psi
_{\tau }=\left( d_{\tau A},d_{\tau B},d_{\tau C},d_{\tau D}\right) $, where $%
d_{\tau M}$ denotes the annihilation operator of electron for orbital $\tau $
at site $M$ with $\tau =1$, $2$, $3$ standing for $d_{\overline{x}\overline{z%
}}$, $d_{\overline{y}\overline{z}}$, $d_{\overline{x}\overline{y}}$,
respectively. We will use $x$ and $y$ as the coordinates of this system and $%
\overline{x}\ $and $\overline{y}$ as the nearest Fe-Fe directions.

\begin{figure}[ptb]
\begin{center}
\includegraphics[
trim=0.252092in 0.000000in 0.317148in 0.000000in, height=3.6668in,
width=3.6045in ] {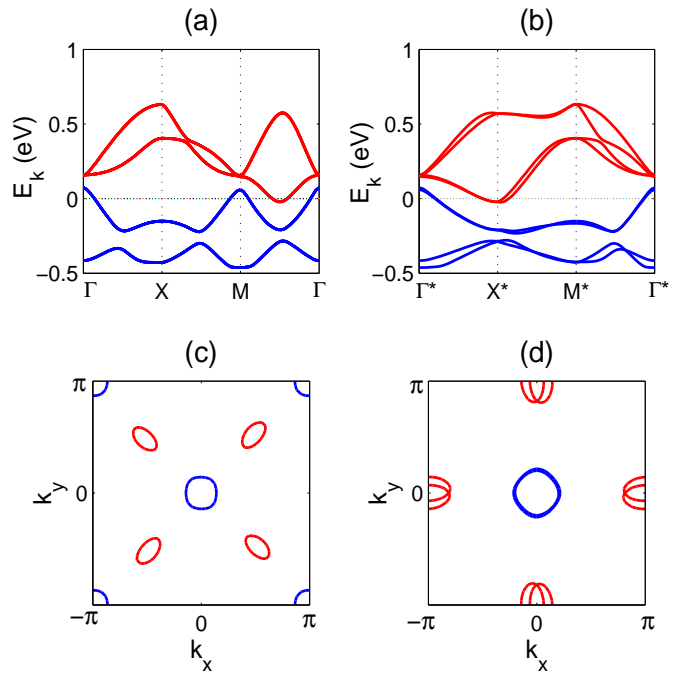}
\end{center}
\caption{(Color online) The band structure of the three--orbtial model with
the $\protect\sqrt {5}\times\protect\sqrt{5}$ lattice structure in the
un-folded BZ (a) and in the folded BZ (b), and its correspoinding Fermi
surfaces in (c) and (d), respectively. We have shifted the dispersion ($%
\protect\mu$=0.557 in our model), such that \textit{E}=0 corresponds to the
Fermi level. The bands shown are only those near the Fermi level, from the
eighth to the eleventh (from low energy to high energy). }
\label{Ek}
\end{figure}

In previous works, the DFT band structure of the parent compound KFe$_{2}$Se$%
_{2}$ in which only electron pockets appear at M \cite%
{TAMaier,FaWang,YZhou,TDas1,Mazin} is employed. However, the parent compound
KFe$_{2}$Se$_{2}$\ does not have the same hopping parameters as those in the
vacancy-ordered K$_{y}$Fe$_{1.6}$Se$_{2}$. For example, there is no hopping
between Fe atoms and vacancies. Here we shall strictly enforce the $\sqrt{5}%
\times \sqrt{5}$\ vacancy-ordered structure and construct a tight-binding
model with nearest neighbor (NN) and next-nearest neighbor (NNN) Fe-Fe
hoppings, which will be classified as intra- and inter-cell ones. As the
vacancy ordering appears, the reflection symmetry is lost but the
four-fold-rotational symmetry is left intact. The system is invariant under
90${}^{\circ }$\textbf{\ }rotations around the center of a unit cell, which
is the position of Se (if the vacancy position is taken as the rotation
center, the rotation has to be followed by a $P_{z}$ operation: the
reflection $z\rightarrow -z$). A general hopping Hamiltonian $H_{t}$ with
vacancy order being included can be written down by imposing the 90${%
{}^{\circ }}$ right-handed rotation symmetry: $d_{1}\rightarrow d_{2}$, $%
d_{2}\rightarrow -d_{1}$, $d_{3}\rightarrow -d_{3}$ accompanied with $%
A\rightarrow B$, $B\rightarrow C$, $C\rightarrow D$, $D\rightarrow A$. Due
to its massive form, the general form of $H_{t}$ is relayed to the Appendix.
After Fourier transformation, in the momentum $k$ space, it takes the form 
\begin{equation}
H_{t}=\sum_{\mathbf{k}}\psi ^{\dag }(\mathbf{k})M(\mathbf{k})\psi (\mathbf{k}%
),
\end{equation}%
where $\psi (\mathbf{k})=(\psi _{1}(\mathbf{k}),\psi _{2}(\mathbf{k}),\psi
_{3}(\mathbf{k}))$ with $\psi _{\tau }(\mathbf{k})=(d_{\tau A,\mathbf{k}%
},d_{\tau B,\mathbf{k}},d_{\tau C,\mathbf{k}},d_{\tau D,\mathbf{k}})$ and $M(%
\mathbf{k})$ is a $12\times 12$ matrix. Detailed characterization of all
hopping parameters are tabulated in TABLE \ref{intra_t} and TABLE \ref%
{inter_t} in Appendix. These parameters are obtained by fitting energy
dispersions (in the folded BZ) to the results of X. W. Yan \textit{et al.} 
\cite{XWYan} obtained by the generalized gradient approximation (GGA) in
which the main features are four hole pockets at $\Gamma $ and four electron
pockets at X in the nonmagnetic state. Our fitting gives two hole and two
electron pockets, which capture basic features of this system. Fig. \ref{Ek}
shows our fitting results in the unfolded BZ ((a) and (c)), and in the
folded BZ ((b) and (d)). In the folded BZ, there are two hole pockets around
($0,0$) and two electron pockets around ($\pi ,0$); in the unfolded
coordinate, one hole pocket will move to ($\pi ,\pi $) and electron pockets
to $\pm $($\pi /2,\pm \pi /2$).

\begin{table}[tb]
\centering
\begin{tabular}{|c||c|c|c|}
\hline
& $n_{A}=n_{C}$ & $n_{B}=n_{D}$ & $1/4\sum_{I}n_{I}$ \\ \hline\hline
$n_{1}$ & 1.92 & 1.74 & 1.83 \\ \hline
$n_{2}$ & 1.74 & 1.92 & 1.83 \\ \hline
$n_{3}$ & 0.84 & 0.84 & 0.84 \\ \hline
$\sum_{\tau}n_{\tau}$ & 4.50 & 4.50 & 4.50 \\ \hline
\end{tabular}%
\caption{Particle number per Fe for different orbitals ($\protect\tau=1, 2,
3 $) and sites ($I=A, B, C, D$). Due to the four-fold rotation symmetry,
some numbers are equal.}
\label{particle_no}
\end{table}

As for the particle number, the stoichiometric compound $%
A_{0.8}Fe_{1.6}Se_{2}$ gives Fe$^{2+}$. In other words, there are six
electrons for each iron. Previous three-band model works for iron-pnictides
claimed four electrons per Fe in the undoped state. 

\bigskip In the GGA calculation \cite{XWYan}, the number of electrons
enclosed by Fermi surfaces is about 0.642 electrons/cell, while the number
of holes enclosed by Fermi surfaces is about 0.529 holes/cell. In our model
at the symmetry point, the electron number per iron is 4.5 and hence the
total number of electrons is 18 per cell. The particle density for each
orbital and site is listed in Table \ref{particle_no}. As we expect, due to
symmetry, $n_{1}$ at site $A$ or $C$ ($B$ or $D$) is the same as $n_{2}$ at
site $B$ or $D$ ($A$ or $C$), while $n_{3}$ is uniform at every site. In
addition, we found that there are about 0.52 electrons/cell and 0.52
holes/cell enclosed by Fermi surfaces. These numbers are close to those
found in the GGA calculation. We note in passing that it is possible to
change the chemical potential and hopping scales so that the model is away
from the symmetry point and numbers of electrons/holes per cell are closer
to those obtained by the GGA calculation. However, since we do not find
significant changes of magnetic properties, we shall be focusing on the
symmetry point.

\section{Magnetic and Charge Responses}

Using the tight-binding model with the fitted parameters found in the last
section, we can analyze linear responses of the system. We shall first
calculate the generalized susceptibility in the absence of the
electron-electron interaction defined by 
\begin{align}
\chi_{0}^{ab,cd}(\mathbf{q},i\Omega_{n}) & =\int d\tau e^{i\Omega_{n}\tau
}\left\langle S_{ab}^{+}(\mathbf{q},\tau)S_{cd}^{-}(-\mathbf{q}%
,0)\right\rangle _{0} \\
& =-\frac{1}{\beta N}\underset{\mathbf{k},i\omega_{n}}{{\textstyle\sum} }%
g_{ca}(\mathbf{k},i\omega_{n})g_{bd}(\mathbf{k+q},i\omega_{n}+i\Omega _{n}).
\notag
\end{align}
Here the generalized spin operators are defined by $S_{ab}^{+} \equiv
\psi_{a,\uparrow}^{\dag}\psi _{b,\downarrow}$ and $S_{cd}^{-} \equiv \left(
S_{cd}^{+}\right) ^{\dag}=\psi_{d,\downarrow}^{\dag}\psi_{c,\uparrow}$ with
the subscript ($a,b,c,d$) being the 12 orbital indices for electrons, and
the Green's function $g_{ab}$ is given by 
\begin{equation}
g_{ab}(\mathbf{k},i\omega_{n})=\underset{\mu}{ {\textstyle\sum} }\frac{%
A_{a\mu}(\mathbf{k})A_{b\mu}^{\ast}(\mathbf{k})}{i\omega_{n}-E_{\mu }(%
\mathbf{k})},
\end{equation}
where $\mu$ is the band index and $A_{a\mu}$ is the orbital-band
transformation matrix, $\psi_{a}(\mathbf{k})=\underset{\mu}{{\textstyle\sum} 
}A_{a\mu}(\mathbf{k})\gamma_{\mu}(\mathbf{k})$. By analytic continuity, the
susceptibility becomes%
\begin{align}
\chi_{0}^{ab,cd}(\mathbf{q},\omega) & =-\frac{1}{N}\underset{\mathbf{k}
,\mu,\nu}{ {\textstyle\sum} }A_{c\mu}(\mathbf{k})A_{a\mu}^{\ast}(\mathbf{k}%
)A_{b\nu}(\mathbf{k+q})A_{d\nu }^{\ast}(\mathbf{k+q})  \notag \\
& \times\frac{n_{F}\left[ E_{\mu}(\mathbf{k})\right] -n_{F}\left[ E_{\nu }(%
\mathbf{k+q})\right] }{\omega+E_{\mu}(\mathbf{k})-E_{\nu}(\mathbf{k+q}
)+i\delta}.
\end{align}
We now include the effect of electron-electron interaction by considering
the generalized Hubbard model, in which all interactions are on the same Fe
atom, 
\begin{align}
H_{I} & =\underset{i}{ {\textstyle\sum} }\underset{I=A,B,C,D}{ {\textstyle%
\sum} }\left\{ U\underset{a=1,2,3}{ {\textstyle\sum} }n_{aI,i%
\uparrow}n_{aI,i\downarrow}\right.  \notag \\
& +\underset{a,b(a>b)}{ {\textstyle\sum} }\left[ \left( U^{\prime}-\frac{%
J_{H}}{2}\right) n_{aI,i}n_{bI,i}-2J_{H}\mathbf{S}_{aI,i}\cdot\mathbf{S}%
_{bI,i}\right.  \notag \\
& \left. \left. +J_{C}\left( d_{aI,i\uparrow}^{\dag}d_{aI,i\downarrow
}^{\dag}d_{bI,i\downarrow}d_{bI,i\uparrow}+h.c.\right) \right] \right\} .
\label{H_I}
\end{align}
Here we simply use the same set of parameters for every site and orbital.
Within this model, we calculate the random-phase approximation (RPA)
susceptibilities for spin and charge 
\begin{align}
\chi_{s,RPA}(\mathbf{q},\omega) & =\frac{\chi_{0}(\mathbf{q},\omega )}{%
1-\Gamma_{s}\chi_{0}(\mathbf{q},\omega)}, \\
\chi_{c,RPA}(\mathbf{q},\omega) & =\frac{\chi_{0}(\mathbf{q},\omega )}{%
1+\Gamma_{c}\chi_{0}(\mathbf{q},\omega)}.
\end{align}
The vertices for spin sector are $\Gamma_{s}^{\tau\tau,\tau\tau}=U$, $%
\Gamma_{s}^{\tau\tau^{\prime},\tau\tau^{\prime}}=U^{\prime}$, $\Gamma
_{s}^{\tau\tau,\tau^{\prime}\tau^{\prime}}=J_{H}$, $\Gamma_{s}^{\tau
\tau^{\prime},\tau^{\prime}\tau}=J_{C}$, and for charge sector $\Gamma
_{c}^{\tau\tau,\tau\tau}=U$, $\Gamma_{c}^{\tau\tau^{\prime},\tau\tau^{%
\prime}}=-U^{\prime}+2J_{H}$, $\Gamma_{c}^{\tau\tau,\tau^{\prime}\tau^{%
\prime}}=2U^{\prime}-J_{H}$, $\Gamma_{c}^{\tau\tau^{\prime},\tau^{\prime}%
\tau}=J_{C}$, where nonvanishing vertices are only between the same Fe, and $%
\tau$ denotes orbitals and $\tau\neq\tau^{\prime}$. In the following, we
shall take the relations $U^{\prime}=U-2J_{H}$ and $J_{C}=J_{H}$.

\begin{figure}[ptb]
\begin{center}
\includegraphics[
height=2.5711in, width=3.6071in ] {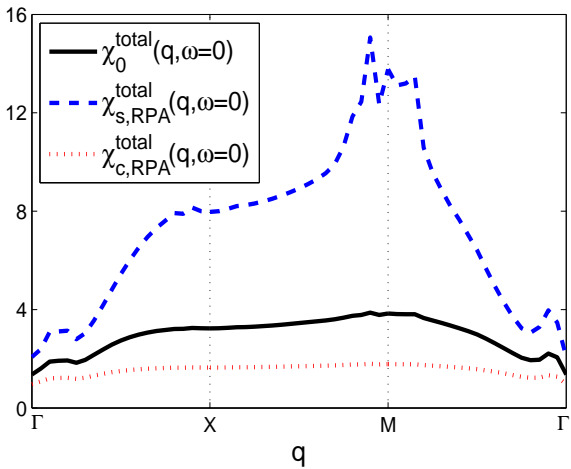}
\end{center}
\caption{(Color online) Total DC susceptibilities. black solid line: bare,
blue dashed line: spin, red dotted line: charge. The interaction parameters
are $U=1.2eV$, $J_{H}=0.2U$, and $U^{\prime}=U-2J_{H}$.}
\label{total_chi}
\end{figure}

In Fig. \ref{total_chi}, we show the total DC susceptibilities per cell
(four iron) defined by $\chi ^{total}(\mathbf{q},0)={\textstyle\sum_{s,t}}%
\chi ^{ss,tt}(\mathbf{q},0)$. Here $U$=1.2eV and $J_{H}=0.2U$ are used. The
black solid line is for the bare susceptibility $\chi _{0}^{total}$, the
blue dashed line is the spin susceptibility, and the red dotted line is for
the charge susceptibility. As expected, electron-electron interaction
strongly enhances spin susceptibility $\chi _{s}^{total}$ and induces a peak
around ($\pi ,\pi $). The Stoner instability for $J_{H}=0.2U$ is found to
happen at $U$=1.5eV and such divergence of $\chi _{s}^{total}$\ at ($\pi
,\pi $) will result in the checkerboard AFM pattern as experiments observed.
Therefore, the fitted tight-binding Hamiltonian explains the experimental
observations. On the other hand, the charge susceptibility $\chi _{c}^{total}
$\ is not important here and is smaller than the bare one, which is
consistent with results of Ref. \cite{Graser2009}. 
\begin{figure*}[tbp]
\centering
{\includegraphics[ height=4.587in, width=7.0378in ] {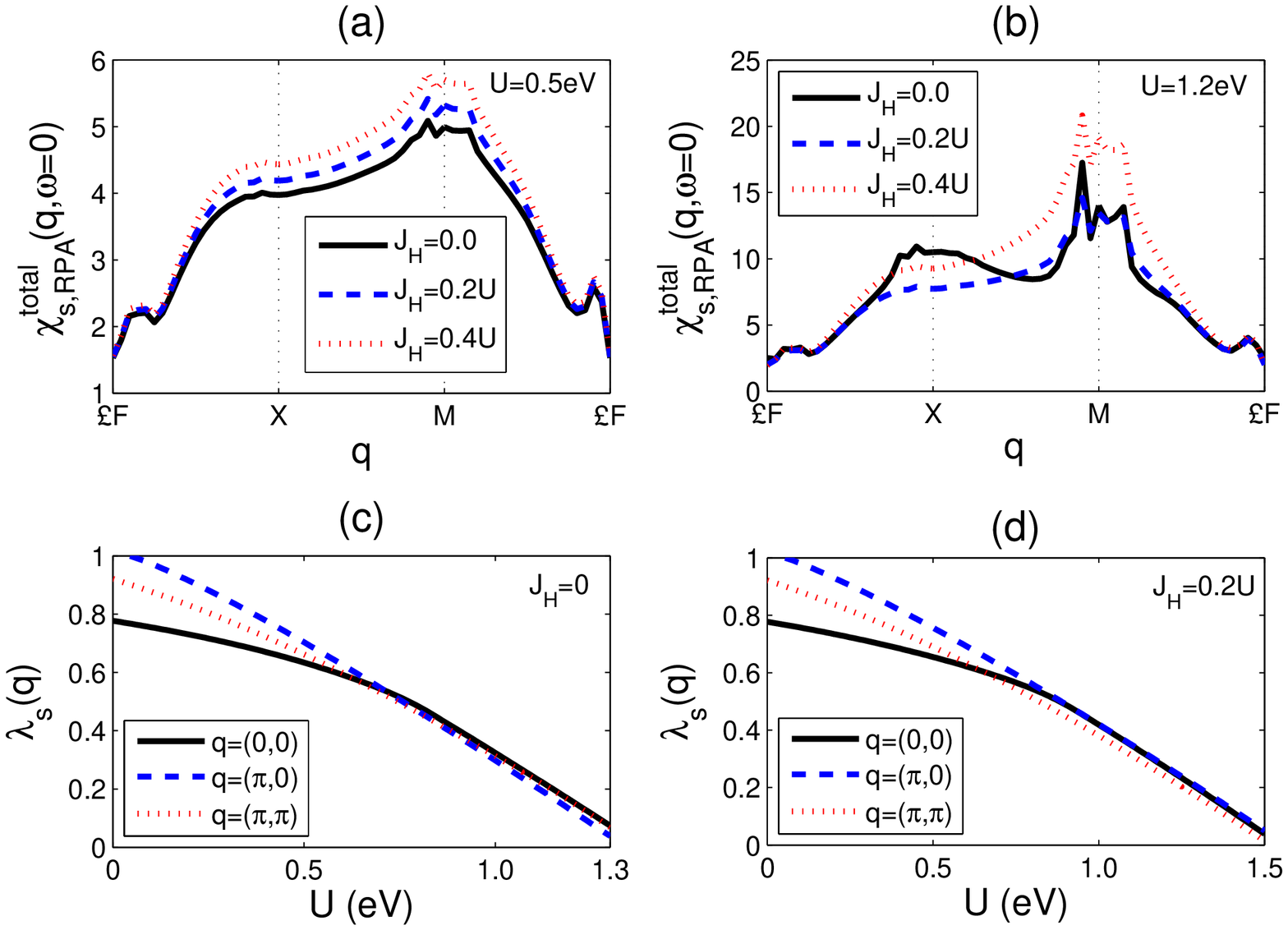} }
\caption{(Color online) Upper two panels: total spin susceptibility $\protect%
\chi _{s,RPA}^{total}(\mathbf{q},0)$ at different $J_{H}$ in $U$=0.5eV (a),
and in $U$=1.2eV (b). Lower two: $\protect\lambda _{s}(\mathbf{q)}$, the
minimal eigenvalue of the inverse of $\protect\chi _{s,RPA}^{total}(\mathbf{q%
},0)$ at three $\mathbf{q}$ vectors in $J_{H}=0$ (c), and in $J_{H}=0.2U$
(d). The magnetic transition happens when $\protect\lambda _{s}(\mathbf{q}%
)=0_{+}$.}
\label{total_xs}
\end{figure*}

Next we investigate the effect of the Hund's-Rule coupling. As shown in Fig. %
\ref{total_xs}(a), for small $U$ (take $U$=0.5eV as a nominal example),
values of $\chi_{s}^{total}$ show a monotonic behavior. However, in the
large $U$ case as shown in Fig. \ref{total_xs}(b) ($U$=1.2eV), values of $%
\chi_{s}^{total}$ exhibit non-monotonic behavior. In particular, the
shoulder around ($\pi,0$) at large $J_{H}$\ becomes a hump at $J_{H}=0$. The
hump at ($\pi,0$) indicates that there is a magnetic instability for striped
AFM at about $U$=1.3eV, in competition with the block-checkerboard AFM.

To further check the magnetic instability, we employ the Stoner criterion.
In the multi-orbital system, the susceptibility is a matrix and magnetic
instability is determined by the corresponding eigenvalues. The Stoner
criterion requires one to find the first eigenvalue, $\lambda _{s}$, that
reaches zero, i.e., $\lambda _{s}(\mathbf{q})=0_{+}$, where $\lambda _{s}(%
\mathbf{q})$ is the minimal eigenvalue of the inverse of $\chi _{s,RPA}(%
\mathbf{q},0)$ [$\chi _{s,RPA}^{-1}(\mathbf{q},0)=\chi _{0}^{-1}(\mathbf{q}%
,\omega )-\Gamma _{s}$] and $\mathbf{q}$ will be the magnetic ordering
vector. Fig. \ref{total_xs}(c) and \ref{total_xs}(d) show the behavior of $%
\lambda _{s}(\mathbf{q})$\ versus $U$. It is seen that at $J_{H}=0$, shown
in Fig. \ref{total_xs}(c), the first eigenvalue that touches zero occurs at $%
\mathbf{q=}$($\pi ,0$). Hence the stripe AFM is the resulting magnetic phase
at $J_{H}=0$, in consistent with our previous conclusion. At larger $J_{H}$,
as shown in Fig. \ref{total_xs}(d) ($J_{H}=0.2U$), the magnetic instability
occurs at $\mathbf{q=}$($\pi ,\pi $). From Fig. \ref{total_xs}(c) and \ref%
{total_xs}(d), we also find that the critical value of $U$, $U_{c}$, when
magnetic instability occurs, depends on $J_{H}$ as well. For $J_{H}=0$, we
find that $U_{c}\sim $1.3eV, while for $J_{H}=0.2U$, we get $U_{c}\sim $%
1.5eV. These results all suggest that large Hund's rule coupling stabilizes
the checkerboard AFM state.

We note in passing that in the above, we do not try to distinguish whether
the magnetic instability occurs exactly at ($\pi ,\pi $) or not. All of
these magnetic states are classified as the checkerboard AFM state. In fact,
because the parameters adopted in Fig. \ref{total_xs} are for the system at
the symmetry point, the magnetic instability does not happen exactly at ($%
\pi ,\pi $). By changing the chemical potential, the wave vector of the
magnetic instability can be shifted to be exactly at ($\pi ,\pi $). This
implies that the exact wave vector for the magnetic instability will
generally depend on the doping level of the system.

\section{Summary and discussion}

In summary, in contrast to perturbative treatment of vacancies \cite{TDas2},
we have constructed an effective tight-binding model for the K$_{y}$Fe$_{1.6}
$Se$_{2}$ system by including exact symmetries of the Fe vacancy ordering
structure. The tight-binding model includes three orbitals ($d_{\overline{x}%
\overline{z}}$, $d_{\overline{y}\overline{z}}$, and $d_{\overline{x}%
\overline{y}}$), which are considered to be the most important orbits in
iron-pnictides and iron-chalcogenides. Although this system shows a large
moment \cite{WBao} and could be better described by including some localized
moments, a proper tight-binding band structure is still required since
iron-based superconductors so far are regarded as an intermediate coupling
system instead of being a strong coupling system. For example, recent
experimental findings from thermal transport of K$_{x}$Fe$_{2-y}$Se$_{2}$
indicated it a weakly or intermediately correlated system \cite{KWang}. From
these aspects, it is clear that our model captured the essential low energy
physics: two hole pockets around $\Gamma $\ and two electron pockets around
X and Y in the folded BZ. Furthermore, the constructed band model, when
combined with generalized Hubbard interactions, yields a spin susceptibility
which exhibits both the block-checkerboard antiferromagnetism instability
and the stripe antiferromagnetism instability. In particular, for large
Hund's rule couplings, the block-checkerboard antiferromagnetism wins over
the stripe antiferromagnetism, in agreement with recent observations in
experiments.

While so far in this work we only consider the magnetic instability of the
ternary iron selenide system, our findings also provide some insight into
possible mechanism for superconductivity occurring in this system. In
particular, the strong spin fluctuations at $\mathbf{q}$=($\pi ,\pi $) could
result in inter hole-pocket and inter electron-pocket (in opposite momenta)
scatterings, which may lead to pairing with totally different symmetries of
pairing. In real space, it implies that inter-block fluctuations of spins
might play a similar role in analogous to spin-fluctuations in high-$T_{c}$
cuprates. While our model has not yet accounted for superconductivity
observed in this system, the fitted tight-binding model shall serve as a
useful starting point for developing the correct theory.

\begin{acknowledgments}
We thank Prof. Ting-Kuo Lee for discussions. This work was supported by the
National Science Council of Taiwan.
\end{acknowledgments}

\appendix

\section{The effective tight-binding Hamiltonian}

In this appendix, we will include details for construction of the tight
binding Hamiltonian. Following the symmetry argument given in the context
and neglect the tetramer lattice distortion \cite{XWYan}, the tight-binding
Hamiltonian with NN and NNN hoppings can be written as 
\begin{equation}
H_{t}=H_{\epsilon}+H_{12}+H_{3}+H_{12,3}.
\end{equation}
Here $H_{\epsilon}$ is the on-site energy. $H_{12}$ characterizes hopping
among orbitals: $d_{\overline{x}\overline{z}}$ and $d_{\overline{y}\overline{%
z}}$, while $H_3$ is the hopping term for $d_{\overline{x}\overline{y}}$ and 
$H_{12,3}$ describes the hopping between $d_{\overline{x}\overline{z}}$/ $d_{%
\overline{y}\overline{z}}$ and $d_{\overline{x}\overline{y}}$.

We shall suppress the spin index for simplicity. To consider the effect of
Se atoms above and below the Fe plane periodically, the transformation $%
d_{3I,i}\rightarrow (-1)^{\left\vert i\right\vert }d_{3I,i}$ is included
implicitly to make the hopping integrals site-independent. Due to symmetries
imposed by the $\sqrt{5}\times \sqrt{5}$ vacancy ordered structure, we find
that the on-site energies for $d_{\overline{x}\overline{z}}$ and $d_{%
\overline{y}\overline{z}}$ are different and their difference will be
denoted by $\Delta $, while the on-site energy of $d_{\overline{x}\overline{y%
}}$ will be denoted by $\epsilon $. The on-site energy can be written as 
\begin{widetext}
\begin{align}
H_{\epsilon}=\frac{\Delta}{2}\sum_{i}\left[  d_{1A,i}^{\dag}d_{1A,i}
+d_{2B,i}^{\dag}d_{2B,i}+d_{1C,i}^{\dag}d_{1C,i}+d_{2D,i}^{\dag}
d_{2D,i}\right. \nonumber\\
\left.  -(1\longleftrightarrow2)\right] \\
+\epsilon\sum_{i}\left[  d_{3A,i}^{\dag}d_{3A,i}+d_{3B,i}^{\dag}
d_{3B,i}+d_{3C,i}^{\dag}d_{3C,i}+d_{3D,i}^{\dag}d_{3D,i}\right]
\nonumber \end{align}
\end{widetext}

To describe hopping terms, we will adopt the notation $t_{mn,\overline{R}}$
for intra-cell hoppings and $t_{mn,\overline{R}} ^{\prime}$ for inter-cell
hoppings. The subscript $mn$ are the orbital indices and $\overline{R}$ is
the Fe-Fe direction. We note that because of the absence of reflection
symmetry, the NN Fe-Fe hopping between $d_{\overline{x}\overline {z}}$ and $%
d_{\overline{y}\overline{z}}$\ is allowable now. By including all possible
terms allowed by symmetries, hopping terms can be generally expressed as 
\begin{widetext}
\begin{align}
H_{12}  &  =\sum_{i}\left\{  t_{11,\overline{x}}\left(  d_{1A,i}^{\dag
}d_{1B,i}+d_{1C,i}^{\dag}d_{1D,i}+d_{2B,i}^{\dag}d_{2C,i}+d_{2D,i}^{\dag
}d_{2A,i}\right)  \right. \\
&  +t_{11,\overline{y}}\left(  d_{1D,i}^{\dag}d_{1A,i}+d_{1B,i}^{\dag}%
d_{1C,i}+d_{2A,i}^{\dag}d_{2B,i}+d_{2C,i}^{\dag}d_{2D,i}\right) \nonumber\\
&  +t_{11,\overline{x}}^{\prime}\left(  d_{1B,i}^{\dag}d_{1D,i-x}%
+d_{2A,i}^{\dag}d_{2C,i+y}\right)  +t_{11,\overline{y}}^{\prime}\left(
d_{1A,i}^{\dag}d_{1C,i+y}+d_{2B,i}^{\dag}d_{2D,i-x}\right) \nonumber\\
&  +t_{11,\overline{x}+\overline{y}}\left(  d_{1A,i}^{\dag}d_{1C,i}%
+d_{2B,i}^{\dag}d_{2D,i}\right)  +t_{11,\overline{x}-\overline{y}}\left(
d_{1B,i}^{\dag}d_{1D,i}+d_{2A,i}^{\dag}d_{2C,i}\right) \nonumber\\
&  +t_{11,\overline{x}+\overline{y}}^{\prime}\left(  d_{1A,i}^{\dag}%
d_{1D,i+y}+d_{1C,i}^{\dag}d_{1B,i-y}+d_{2B,i}^{\dag}d_{2A,i-x}+d_{2D,i}^{\dag
}d_{2C,i+x}\right) \nonumber\\
&  +t_{11,\overline{x}-\overline{y}}^{\prime}\left(  d_{1A,i}^{\dag}%
d_{1B,i+x}+d_{1C,i}^{\dag}d_{1D,i-x}+d_{2B,i}^{\dag}d_{2C,i+y}+d_{2D,i}^{\dag
}d_{2A,i-y}\right) \nonumber\\
&  +t_{12,\overline{x}}\left(  d_{1A,i}^{\dag}d_{2B,i}+d_{1C,i}^{\dag}%
d_{2D,i}-d_{2B,i}^{\dag}d_{1C,i}-d_{2D,i}^{\dag}d_{1A,i}\right) \nonumber\\
&  +t_{12,\overline{y}}\left(  d_{2A,i}^{\dag}d_{1B,i}+d_{2C,i}^{\dag}%
d_{1D,i}-d_{1B,i}^{\dag}d_{2C,i}-d_{1D,i}^{\dag}d_{2A,i}\right) \nonumber\\
&  +t_{12,\overline{x}}^{\prime}\left(  d_{1B,i}^{\dag}d_{2D,i-x}%
+d_{1D,i}^{\dag}d_{2B,i+x}-d_{1A,i}^{\dag}d_{1C,i+y}-d_{2C,i}^{\dag}%
d_{1A,i-y}\right) \nonumber\\
&  +t_{12,\overline{x}+\overline{y}}\left(  d_{1A,i}^{\dag}d_{2C,i}%
+d_{1C,i}^{\dag}d_{2A,i}-d_{2B,i}^{\dag}d_{1D,i}-d_{2D,i}^{\dag}%
d_{1B,i}\right) \nonumber\\
&  +t_{12,\overline{x}+\overline{y}}^{\prime}\left(  d_{1A,i}^{\dag}%
d_{2D,i+y}+d_{1C,i}^{\dag}d_{2B,i-y}-d_{2B,i}^{\dag}d_{1A,i-x}-d_{2D,i}^{\dag
}d_{1C,i+x}\right) \nonumber\\
&  \left.  +t_{12,\overline{x}-\overline{y}}^{\prime}\left(  d_{2A,i}^{\dag
}d_{1D,i+y}+d_{2C,i}^{\dag}d_{1B,i-y}-d_{1B,i}^{\dag}d_{2A,i-x}-d_{1D,i}%
^{\dag}d_{2C,i+x}\right)  +h.c.\right\}, \nonumber
\end{align}
\begin{align}
H_{3}  &  =\sum_{i}\left\{  t_{33,\overline{x}}\left(  d_{3A,i}^{\dag}%
d_{3B,i}+d_{3B,i}^{\dag}d_{3C,i}+d_{3C,i}^{\dag}d_{3D,i}+d_{3D,i}^{\dag
}d_{3A,i}\right)  \right. \\
&  t_{33,\overline{x}}^{\prime}\left(  d_{3A,i}^{\dag}d_{3C,i+y}%
+d_{3B,i}^{\dag}d_{3D,i-x}\right)  +t_{33,\overline{x}+\overline{y}}\left(
d_{3A,i}^{\dag}d_{3C,i}+d_{3B,i}^{\dag}d_{3D,i}\right) \nonumber\\
&  \left.  +t_{33,\overline{x}+\overline{y}}^{\prime}\left(  d_{3A,i}^{\dag
}d_{3D,i+y}+d_{3B,i}^{\dag}d_{3A,i-x}+d_{3C,i}^{\dag}d_{3B,i-y}+d_{3D,i}%
^{\dag}d_{3C,i+x}\right)  +h.c.\right\}, \nonumber
\end{align}
\begin{align}
{\rm and}\, H_{12,3}  &  =\sum_{i}\left\{  t_{13,\overline{x}}\left(
d_{3A,i}^{\dag
}d_{1B,i}-d_{3B,i}^{\dag}d_{2C,i}-d_{3C,i}^{\dag}d_{1D,i}+d_{3D,i}^{\dag
}d_{2A,i}\right)  \right. \\
&  +t_{13,\overline{y}}\left(  d_{3A,i}^{\dag}d_{1D,i}-d_{3B,i}^{\dag}%
d_{2A,i}-d_{3C,i}^{\dag}d_{1B,i}+d_{3D,i}^{\dag}d_{2C,i}\right) \nonumber\\
&  +t_{23,\overline{x}}\left(  d_{1A,i}^{\dag}d_{3D,i}-d_{2B,i}^{\dag}%
d_{3A,i}-d_{1C,i}^{\dag}d_{3B,i}+d_{1D,i}^{\dag}d_{3C,i}\right) \nonumber\\
&  +t_{23,\overline{y}}\left(  d_{1A,i}^{\dag}d_{3B,i}-d_{2B,i}^{\dag}%
d_{3C,i}-d_{1C,i}^{\dag}d_{3D,i}+d_{2D,i}^{\dag}d_{3A,i}\right) \nonumber\\
&  +t_{13,\overline{x}}^{\prime}\left(  d_{1D,i}^{\dag}d_{3B,i+x}%
-d_{2A,i}^{\dag}d_{3C,i+y}-d_{1B,i}^{\dag}d_{3D,i-x}+d_{2C,i}^{\dag}%
d_{3A,i-y}\right) \nonumber\\
&  +t_{13,\overline{y}}^{\prime}\left(  d_{1A,i}^{\dag}d_{3C,i+y}%
-d_{2B,i}^{\dag}d_{3D,i-x}-d_{1C,i}^{\dag}d_{3A,i-y}+d_{2D,i}^{\dag}%
d_{3B,i+x}\right) \nonumber\\
&  +t_{13,\overline{x}+\overline{y}}\left(  d_{1C,i}^{\dag}d_{3A,i}%
-d_{2D,i}^{\dag}d_{3B,i}-d_{1A,i}^{\dag}d_{3C,i}+d_{2B,i}^{\dag}%
d_{3D,i}\right) \nonumber\\
&  +t_{13,\overline{x}-\overline{y}}\left(  d_{1B,i}^{\dag}d_{3D,i}%
-d_{2C,i}^{\dag}d_{3A,i}-d_{1D,i}^{\dag}d_{3B,i}+d_{2A,i}^{\dag}%
d_{3C,i}\right) \nonumber\\
&  +t_{13,\overline{x}+\overline{y}}^{\prime}\left(  d_{1A,i}^{\dag}%
d_{3D,i+y}-d_{2B,i}^{\dag}d_{3A,i-x}-d_{1C,i}^{\dag}d_{3B,i-y}+d_{2D,i}^{\dag
}d_{3C,i+x}\right) \nonumber\\
&  +t_{13,\overline{x}-\overline{y}}^{\prime}\left(  d_{1A,i}^{\dag}%
d_{3B,i+x}-d_{1C,i}^{\dag}d_{3D,i-x}-d_{2B,i}^{\dag}d_{3C,i+y}+d_{2D,i}^{\dag
}d_{3A,i-y}\right) \nonumber\\
&  +t_{23,\overline{x}+\overline{y}}^{\prime}\left(  d_{2A,i}^{\dag}%
d_{3D,i+y}-d_{2C,i}^{\dag}d_{3B,i-y}+d_{1B,i}^{\dag}d_{3A,i-x}-d_{1D,i}^{\dag
}d_{3C,i+x}\right) \nonumber\\
&  \left.  +t_{23,\overline{x}-\overline{y}}^{\prime}\left(  d_{2A,i}^{\dag
}d_{3B,i+x}-d_{2C,i}^{\dag}d_{3D,i-x}+d_{1B,i}^{\dag}d_{3C,i+y}-d_{1D,i}%
^{\dag}d_{3A,i-y}\right)  +h.c.\right\}  .\nonumber
\end{align}
\end{widetext}

After Fourier transformation, the Hamiltonian is written in a matrix form as 
\begin{equation}
H_{t}=\sum_{\mathbf{k}}\psi ^{\dag }(\mathbf{k})M(\mathbf{k})\psi (\mathbf{k}%
)
\end{equation}%
where the basis vector is defined as before, $\psi (\mathbf{k})=(\psi _{1}(%
\mathbf{k}),\psi _{2}(\mathbf{k}),\psi _{3}(\mathbf{k}))$ with $\psi _{\tau
}(\mathbf{k})=(d_{\tau A,\mathbf{k}},d_{\tau B,\mathbf{k}},d_{\tau C,\mathbf{%
k}},d_{\tau D,\mathbf{k}})$ and the $12\times 12$ matrix $M(\mathbf{k})$\ is
given by 
\begin{equation}
M(\mathbf{k})=\left[ 
\begin{array}{ccc}
M_{11}(\mathbf{k}) & M_{12}(\mathbf{k}) & M_{13}(\mathbf{k}) \\ 
M_{12}^{\dag }(\mathbf{k}) & M_{22}(\mathbf{k}) & M_{23}(\mathbf{k}) \\ 
M_{13}^{\dag }(\mathbf{k}) & M_{23}^{\dag }(\mathbf{k}) & M_{33}(\mathbf{k})%
\end{array}%
\right]
\end{equation}%
with elements being give by 
\begin{widetext}
\begin{equation}
M_{11}(\mathbf{k})=\left[
\begin{array}
[c]{cccc} \frac{\Delta}{2} &
t_{11,\overline{x}}+t_{11,\overline{x}-\overline{y}
}^{\prime}e^{ik_{x}} &
t_{11,\overline{x}+\overline{y}}+t_{11,\overline{y}
}^{\prime}e^{ik_{y}} &
t_{11,\overline{y}}+t_{11,\overline{x}+\overline{y}
}^{\prime}e^{ik_{y}}\\
t_{11,\overline{x}}+t_{11,\overline{x}-\overline{y}
}^{\prime}e^{-ik_{x}} & -\frac{\Delta}{2} &
t_{11,\overline{y}}+t_{11,\overline{x}+\overline{y}
}^{\prime}e^{ik_{y}} &
t_{11,\overline{x}-\overline{y}}+t_{11,\overline{x}
}^{\prime}e^{-ik_{x}}\\
t_{11,\overline{x}+\overline{y}}+t_{11,\overline{y}
}^{\prime}e^{-ik_{y}}&
t_{11,\overline{y}}+t_{11,\overline{x}+\overline{y}
}^{\prime}e^{-ik_{y}} & \frac{\Delta}{2} &
t_{11,\overline{x}}+t_{11,\overline{x}-\overline{y}
}^{\prime}e^{-ik_{x}}\\
t_{11,\overline{y}}+t_{11,\overline{x}+\overline{y}
}^{\prime}e^{-ik_{y}}&
t_{11,\overline{x}-\overline{y}}+t_{11,\overline{x}
}^{\prime}e^{ik_{x}} &
t_{11,\overline{x}}+t_{11,\overline{x}-\overline{y}
}^{\prime}e^{ik_{x}}& -\frac{\Delta}{2}
\end{array}
\right],
\end{equation}
\begin{equation}
M_{22}(\mathbf{k})=\left[
\begin{array}
[c]{cccc} -\frac{\Delta}{2} &
t_{11,\overline{y}}+t_{11,\overline{x}+\overline{y}
}^{\prime}e^{ik_{x}} &
t_{11,\overline{x}-\overline{y}}+t_{11,\overline{x}
}^{\prime}e^{ik_{y}} &
t_{11,\overline{x}}+t_{11,\overline{x}-\overline{y}
}^{\prime}e^{ik_{y}}\\
t_{11,\overline{y}}+t_{11,\overline{x}+\overline{y}
}^{\prime}e^{-ik_{x}}& \frac{\Delta}{2} &
t_{11,\overline{x}}+t_{11,\overline{x}-\overline{y}
}^{\prime}e^{ik_{y}} &
t_{11,\overline{x}+\overline{y}}+t_{11,\overline{y}
}^{\prime}e^{-ik_{x}}\\
t_{11,\overline{x}-\overline{y}}+t_{11,\overline{x}
}^{\prime}e^{-ik_{y}}&
t_{11,\overline{x}}+t_{11,\overline{x}-\overline{y}
}^{\prime}e^{-ik_{y}} & -\frac{\Delta}{2} &
t_{11,\overline{y}}+t_{11,\overline{x}+\overline{y}
}^{\prime}e^{-ik_{x}}\\
t_{11,\overline{x}}+t_{11,\overline{x}-\overline{y}
}^{\prime}e^{-ik_{y}}&
t_{11,\overline{x}+\overline{y}}+t_{11,\overline{y}
}^{\prime}e^{ik_{x}} &
t_{11,\overline{y}}+t_{11,\overline{x}+\overline{y}
}^{\prime}e^{ik_{x}} & \frac{\Delta}{2}
\end{array}
\right],
\end{equation}
\begin{equation}
M_{33}(\mathbf{k})=\left[
\begin{array}
[c]{cccc} \epsilon &
t_{33,\overline{x}}+t_{33,\overline{x}+\overline{y}}^{\prime
}e^{ik_{x}} &
t_{33,\overline{x}+\overline{y}}+t_{33,\overline{x}}^{\prime
}e^{ik_{y}} &
t_{33,\overline{x}}+t_{33,\overline{x}+\overline{y}}^{\prime
}e^{ik_{y}}\\
t_{33,\overline{x}}+t_{33,\overline{x}+\overline{y}}^{\prime
}e^{-ik_{x}}& \epsilon &
t_{33,\overline{x}}+t_{33,\overline{x}+\overline{y}}^{\prime
}e^{ik_{y}} &
t_{33,\overline{x}+\overline{y}}+t_{33,\overline{x}}^{\prime
}e^{-ik_{x}}\\
t_{33,\overline{x}+\overline{y}}+t_{33,\overline{x}}^{\prime
}e^{-ik_{y}}&
t_{33,\overline{x}}+t_{33,\overline{x}+\overline{y}}^{\prime
}e^{-ik_{y}}  & \epsilon &
t_{33,\overline{x}}+t_{33,\overline{x}+\overline{y}}^{\prime
}e^{-ik_{x}}\\
t_{33,\overline{x}}+t_{33,\overline{x}+\overline{y}}^{\prime
}e^{-ik_{y}}&
t_{33,\overline{x}+\overline{y}}+t_{33,\overline{x}}^{\prime
}e^{ik_{x}} &
t_{33,\overline{x}}+t_{33,\overline{x}+\overline{y}}^{\prime
}e^{ik_{x}} & \epsilon
\end{array}
\right],
\end{equation}%
\begin{equation}
M_{12}(\mathbf{k})=\left[
\begin{array}
[c]{cccc}%
0 & t_{12,\overline{x}}-t_{12,\overline{x}+\overline{y}}^{\prime}e^{ik_{x}} &
t_{12,\overline{x}+\overline{y}}-t_{12,\overline{x}}^{\prime}e^{ik_{y}} &
-t_{12,\overline{x}}+t_{12,\overline{x}+\overline{y}}^{\prime}e^{ik_{y}}\\
t_{12,\overline{y}}-t_{12,\overline{x}-\overline{y}}^{\prime}e^{-ik_{x}} & 0 &
-t_{12,\overline{y}}+t_{12,\overline{x}-\overline{y}}^{\prime}e^{ik_{y}} &
-t_{12,\overline{x}+\overline{y}}+t_{12,\overline{x}}^{\prime}e^{-ik_{x}}\\
t_{12,\overline{x}+\overline{y}}-t_{12,\overline{x}}^{\prime}e^{-ik_{y}} &
-t_{12,\overline{x}}+t_{12,\overline{x}+\overline{y}}^{\prime}e^{-ik_{y}} &
0 & t_{12,\overline{x}}-t_{12,\overline{x}+\overline{y}}^{\prime}e^{-ik_{x}}\\
-t_{12,\overline{y}}+t_{12,\overline{x}-\overline{y}}^{\prime}e^{-ik_{y}} &
-t_{12,\overline{x}+\overline{y}}+t_{12,\overline{x}}^{\prime}e^{ik_{x}} &
t_{12,\overline{y}}-t_{12,\overline{x}-\overline{y}}^{\prime}e^{ik_{x}} & 0
\end{array}
\right],
\end{equation}
\begin{equation}
M_{13}(\mathbf{k})=\left[
\begin{array}
[c]{cccc}%
0 & t_{23,\overline{y}}+t_{13,\overline{x}-\overline{y}}^{\prime}e^{ik_{x}} &
-t_{13,\overline{x}+\overline{y}}+t_{13,\overline{y}}^{\prime}e^{ik_{y}} &
t_{23,\overline{x}}+t_{13,\overline{x}+\overline{y}}^{\prime}e^{ik_{y}}\\
t_{13,\overline{x}}+t_{23,\overline{x}+\overline{y}}^{\prime}e^{-ik_{x}} & 0 &
-t_{13,\overline{y}}+t_{23,\overline{x}-\overline{y}}^{\prime}e^{ik_{y}} &
t_{13,\overline{x}-\overline{y}}-t_{13,\overline{x}}^{\prime}e^{-ik_{x}}\\
t_{13,\overline{x}+\overline{y}}-t_{13,\overline{y}}^{\prime}e^{-ik_{y}} &
-t_{23,\overline{x}}-t_{13,\overline{x}+\overline{y}}^{\prime}e^{-ik_{y}} &
0 & -t_{23,\overline{y}}-t_{13,\overline{x}-\overline{y}}^{\prime}e^{-ik_{x}%
}\\
t_{13,\overline{y}}-t_{23,\overline{x}-\overline{y}}^{\prime}e^{-ik_{y}} &
-t_{13,\overline{x}-\overline{y}}+t_{13,\overline{x}}^{\prime}e^{ik_{x}} &
-t_{13,\overline{x}}-t_{23,\overline{x}+\overline{y}}^{\prime}e^{ik_{x}} & 0
\end{array}
\right],
\end{equation}
and
\begin{equation}
M_{23}(\mathbf{k})=\left[
\begin{array}
[c]{cccc}%
0 & -t_{13,\overline{y}}+t_{23,\overline{x}-\overline{y}}^{\prime}e^{ik_{x}} &
t_{13,\overline{x}-\overline{y}}-t_{13,\overline{x}}^{\prime}e^{ik_{y}} &
t_{13,\overline{x}}+t_{23,\overline{x}+\overline{y}}^{\prime}e^{ik_{y}}\\
-t_{23,\overline{x}}-t_{13,\overline{x}+\overline{y}}^{\prime}e^{-ik_{x}} &
0 & -t_{23,\overline{y}}-t_{13,\overline{x}-\overline{y}}^{\prime}e^{ik_{y}} &
t_{13,\overline{x}+\overline{y}}-t_{13,\overline{y}}^{\prime}e^{-ik_{x}}\\
-t_{13,\overline{x}-\overline{y}}+t_{13,\overline{x}}^{\prime}e^{-ik_{y}} &
-t_{13,\overline{x}}-t_{23,\overline{x}+\overline{y}}^{\prime}e^{-ik_{y}} &
0 & t_{13,\overline{y}}-t_{23,\overline{x}-\overline{y}}^{\prime}e^{-ik_{x}}\\
t_{23,\overline{y}}+t_{13,\overline{x}-\overline{y}}^{\prime}e^{-ik_{y}} &
-t_{13,\overline{x}+\overline{y}}+t_{13,\overline{y}}^{\prime}e^{ik_{x}} &
t_{23,\overline{x}}+t_{13,\overline{x}+\overline{y}}^{\prime}e^{ik_{x}} & 0
\end{array}
\right]
\end{equation}
\end{widetext}

Our fitting values are $\Delta$=0.2 and $\epsilon$=0.55, and those of the
hopping integrals are listed in TABLE \ref{intra_t} and TABLE \ref{inter_t}. 
\begin{table}[ptb]
\centering
\begin{tabular}{c|cccc}
\hline
$t_{mn,\overline{R}}$ & $\overline{R}=\overline{x}$ & $\overline{R}=%
\overline{y}$ & $\overline{R}=\overline{x}+\overline{y}$ & $\overline {R}=%
\overline{x}-\overline{y}$ \\ \hline
$mn=11$ & -0.14 & -0.09 & 0.03 & 0.03 \\ 
$mn=33$ & -0.05 &  & 0.3 &  \\ 
$mn=12$ & 0 & 0 & 0 &  \\ 
$mn=13$ & -0.25 & 0 & -0.05 & 0.15 \\ 
$mn=23$ & 0 & -0.1 &  &  \\ \hline
\end{tabular}%
\caption{Fitted intra-cell hopping parameters between NN and NNN.}
\label{intra_t}
\end{table}
\begin{table}[ptb]
\centering
\begin{tabular}{c|cccc}
\hline
$t_{mn,\overline{R}}^{\prime}$ & $\overline{R}=\overline{x}$ & $\overline {R}%
=\overline{y}$ & $\overline{R}=\overline{x}+\overline{y}$ & $\overline {R}=%
\overline{x}-\overline{y}$ \\ \hline
$mn=11$ & -0.028 & -0.04 & 0.024 & 0.024 \\ 
$mn=33$ & -0.05 &  & 0.35 &  \\ 
$mn=12$ & 0 &  & -0.03 & -0.06 \\ 
$mn=13$ & -0.375 & 0 & -0.05 & 0.1 \\ 
$mn=23$ &  &  & 0.15 & 0.05 \\ \hline
\end{tabular}%
\caption{Fitted inter-cell hopping parameters between NN and NNN.}
\label{inter_t}
\end{table}

\end{document}